\documentstyle[preprint, floats, aps]{revtex}

\begin{document}

\draft

\preprint{CALT-68-2100, QUIC-97-004}

\title{A Theory of Fault-Tolerant Quantum Computation}
\author{Daniel Gottesman\thanks{gottesma@theory.caltech.edu}}
\address{California Institute of Technology, Pasadena, CA 91125}
\maketitle

\begin{abstract}
In order to use quantum error-correcting codes to actually improve
the performance of a quantum computer, it is necessary to be able
to perform operations fault-tolerantly on encoded states.  I present
a general theory of fault-tolerant operations based on symmetries of
the code stabilizer.  This allows a straightforward determination of
which operations can be performed fault-tolerantly on a given code.
I demonstrate that fault-tolerant universal computation is possible
for any stabilizer code.  I discuss a number of examples in more detail,
including the five-qubit code.
\end{abstract}


The development of quantum error-correcting codes 
\cite{shor1,calderbank1,steane1,gottesman,calderbank2,calderbank3}
has stirred great hopes for conquering errors and decoherence in
quantum computers.  However, just the existence of codes, even very
good codes, is not sufficient.  It is also necessary to be able to
perform operations on encoded states without a catastrophic spread
of existing errors.  However, until now, it was only known how
to implement a universal set of gates fault-tolerantly on a few 
codes~\cite{shor2,knill3,aharonov}.

While a general formalism for quantum error-correcting codes is known,
the code stabilizer~\cite{gottesman,calderbank2,calderbank3}, the
theory of fault-tolerant operations is in a great deal of disarray.
A quantum gate, unlike a classical gate, can cause errors to spread
both forwards and backwards through the gate.  The goal of fault-tolerant
operations is to prevent the spread of errors within a block, which could
change a single correctable error into two errors, which is perhaps more
than the code could handle.  Even if we use codes that correct more than
one error, the spread of errors within a block rapidly reduces the code's
tolerance for errors.  Therefore, I define a fault-tolerant operation to
be one for which a single operational error can only produce one error
within a single encoded block.  The assumption is that storage errors
on different qubits are independent and that gate errors can only affect
qubits which interact via that gate.

A {\em transversal} operation, in which the 
operation acts independently on each qubit in the block, is a
prototypical fault-tolerant operation.  For instance, a bitwise 
controlled NOT operation from one block to another is fault-tolerant, 
since errors can spread only between corresponding qubits in the two blocks.

Unfortunately, most bitwise operations applied to most codes will not
map one valid codeword to another.  Until now, there has been no general
theory of what a bitwise operation will do to the codewords of a given
code.  I show below that a bitwise operation will transform the stabilizer
of a code.  If the stabilizer is rearranged, but otherwise left unchanged,
the operation will take codewords to codewords.  This will give us a
few basic operations on various codes with which to start our analysis.

In the quest to perform universal quantum computation, we are not limited
to unitary operations.  We can also perform measurements.  In
section~\ref{measurements}, I analyze the behavior of certain states
when a measurement is made.  This allows us to see what operations we
can derive from the basic operations by using ancillas and making partial
measurements of the state.  Ultimately, this will allow us to perform
universal computation on a large set of codes, among them the five-qubit
code, a class of distance 2 codes, and the code encoding 3 qubits in
8 qubits.

\section{Encoded NOT and Phase}
\label{normalizer}

Before I advance into the full theory of fault-tolerant operations, I
will discuss how to perform encoded NOT and Phase gates on any stabilizer
code.  The behavior of these gates under more general transformations will
tell us what those transformations actually do to the encoded states.

The stabilizer $S$ of a code is an abelian subgroup of the group $\cal{G}$
generated by the operations
\begin{equation}
I = \pmatrix{ 1 & \ 0 \cr 0 & \ 1 }, 
\ \ 
X = \pmatrix{ 0 & \ 1 \cr 1 & \ 0 }, 
\ \ 
Z = \pmatrix{ 1 & \ 0 \cr 0 & -1 }, 
\ {\rm and}\ 
Y = X \cdot 
Z = \pmatrix{ 0 & -1 \cr 1 & \ 0 }
\end{equation}
acting on each of the $n$ qubits of the code.  I will sometimes write
${\cal G}_n$ to explicitly indicate the number of qubits acted on.
The codewords of the code
are the states $|\psi \rangle$ which are left fixed by every operator
in $S$.  Operators in $\cal{G}$ which anticommute with some operator
in $S$ will take codewords from the coding space into some orthogonal
space.  By making a measurement to distinguish the various orthogonal
spaces, we can then determine what error has occured and correct it.
A quantum code encoding $k$ qubits in $n$ qubits will have a stabilizer
with $n-k$ generators.

However, there are, in general, a number of operators in $\cal{G}$
that commute with all of the operators in $S$.  The set of such operators
is the {\em normalizer} $N(S)$ of $S$ in $\cal{G}$.\footnote{Strictly 
speaking, this is the centralizer of $S$, but in this
case it is equal to the normalizer, since $G^{-1} M G = \pm G^{-1} G M
= \pm M$, and not both $M$ and $-M$ are in $S$.}
$S$ is itself contained in the normalizer, but in general the normalizer
is larger than just $S$.  If $S$ contains $2^a$ operators (so it has $a$
generators), the normalizer will be generated by $2n-a = n+k$ operators.
($\cal{G}$ has a total of $2^{2n}$ elements, half of which will commute
with any other fixed element of $\cal{G}$.)

The elements of the normalizer will change one codeword to another,
and therefore have a natural interpretation in terms
of encoded operations on the code words.  Suppose we extend the stabilizer
into a maximal set of $n$ commuting operators.  Then consider those codewords
which, besides being $+1$-eigenvectors of the stabilizer generators, are
also eigenvectors of the additional $k$ operators from $N(S)$.  Let these
codewords be the basis codewords for our code, giving the encoded 
$|0\ldots 00 \rangle,\ |0\ldots 01 \rangle,\ \ldots |1\ldots 11 \rangle$.
The state which has eigenvalue $+1$ for the $k$ new operators will be
the encoded $|0 \ldots 0 \rangle$, the state which has eigenvalue $-1$
for everything will be the encoded $|1 \ldots 1 \rangle$, and so on.
Then the $k$ new operators have the interpretation of being the encoded
$Z$ operators on the $k$ encoded qubits.  We will write these encoded
$Z$ operators as $\overline{Z_i}$ for the $i$th encoded qubit, or
$\overline{Z}$ when there is just one encoded qubit.

Now, the remaining elements of $N(S)$ will not commute with all of the
encoded $Z$ operators.  We can choose a set of generators for $N(S)$ such
that each of the last $k$ operators commutes with everything except a
single one of the $\overline{Z}$ operators.  These generators are then
the encoded bit flip operators $\overline{X_i}$ (or $\overline{X}$ when there
is just one).  An arbitrary element of $N(S)$ is some other encoded
operation.  If two elements of $N(S)$ differ by an element of the stabilizer,
they act the same way on any code word (since the stabilizer element just
fixes the codeword).  Therefore the actual set of encoded operations
represented in $N(S)$ is $N(S)/S$.

DiVicenzo and Shor showed how to perform syndrome measurement and error
correction fault-tolerantly on any stabilizer code \cite{divincenzo}.
Using the same methods, we can measure the eigenvalue of any operator
in $\cal{G}$, even if it is not in $S$.  This also enables us to prepare
the encoded zero state of any stabilizer code by performing error correction
and measuring the eigenvalue of the $\overline{Z}$ operators.

\section{More General Operations}

So far, we have only considered applying products of $X$, $Y$, and $Z$ to
the codewords.  However, this is not the most general thing we could do.
Suppose we have some totally arbitrary unitary transformation $U$ we
wish to apply to our codewords.  How does this affect other operators,
such as the elements of $S$ and $N(S)$?
\begin{equation}
U M |\psi \rangle = U M U^{\dagger} U |\psi \rangle,
\end{equation}
so $|\psi \rangle$ is an eigenvector of $M$ if and only if $U |\psi \rangle$
is an eigenvector of $U M U^{\dagger}$.  Furthermore, they have the same
eigenvalue.  Thus, by applying $U$ to $|\psi \rangle$, we effectively
transform any operator $M$ of interest into $U M U^\dagger$.
In order for the state $|\psi \rangle$ to remain a codeword,
the state $U|\psi \rangle$ must still be in the coding space, so
$U M U^{\dagger}$ must also fix all the codewords
$|\psi \rangle$ for every $M \in S$.  Let us consider a restricted set
of possible $U$s, those for which $U M U^\dagger$ is actually in $\cal{G}$
(so U is in the normalizer $N(\cal{G})$ of $\cal{G}$ in $U(n)$).
We will see that $N(\cal{G})$ is generated by Hadamard rotations,
$\pi/2$ phase rotations, and controlled NOT
operations~\cite{calderbank2,bennett}.
Then, by the definition of the stabilizer and the coding space, we
need $U M U^\dagger$ to actually be in $S$ for all $M \in S$.  Therefore,
$U$ is actually in the normalizer of $S$ in $U(n)$.  The same criterion
was found previously by Knill~\cite{knill4}.  Note that the normalizer of
$S$ in $U(n)$ is not necessarily a subset of $N(\cal{G})$.

When we restrict our attention to operations that are in both the 
normalizer of $\cal{G}$ in $U(n)$ and the normalizer of $S$ in $U(n)$, it
becomes straightforward to determine the operation actually performed on the
encoded states. First, note that the $\overline{X}$ and $\overline{Z}$ 
operators transform into operators that also commute with everything
in $S$.  Thus, we can rewrite them as products of the original
$\overline{X}$s, $\overline{Z}$s, and elements of $S$.  The elements
of $S$ just give us the equivalence between elements of $N(S)$
discussed in section~\ref{normalizer}, so we have deduced a transformation
of the encoded $X$ and $Z$ operators.  Furthermore, we know this
encoded transformation also lies in the normalizer of ${\cal G}_k$.  

Typically, we want to consider transversal operations $U$, which
are equal to the tensor product of single-qubit operations (or 
operations that only affect one qubit per block).  For the moment,
we will only consider operations of this form and see what collections of
them will do to the stabilizer. Before launching into an analysis of
which gates can be used on which codes, I will present an overview of
the gates that are amenable to this sort of analysis.

For instance, one of the simplest and most common fault-tolerant operations
is the Hadamard rotation
\begin{equation}
R = \frac{1}{\sqrt{2}} \pmatrix{ 1 & \ 1 \cr 1 & -1}.
\end{equation}
Let us see what this does to $X$, $Y$, and $Z$.
\begin{eqnarray}
R X R^\dagger = \frac{1}{2} \pmatrix{ 1 & \ 1 \cr 1 & -1}
\pmatrix{ 1 & -1 \cr 1 & \ 1} = \pmatrix{ 1 & \ 0 \cr 0 & -1} = & Z \\
R Z R^\dagger = \frac{1}{2} \pmatrix{ 1 & \ 1 \cr 1 & -1}
\pmatrix{ \ 1 & 1 \cr -1 & 1} = \pmatrix{ 0 & 1 \cr 1 & 0} = & X \\
R Y R^\dagger = \frac{1}{2} \pmatrix{ 1 & \ 1 \cr 1 & -1}
\pmatrix{ -1 & 1 \cr \ 1 & 1} = \pmatrix{ \ 0 & 1 \cr -1 & 0} = & -Y.
\end{eqnarray}
Therefore, applying $R$ bitwise will switch all the $X$s and all the $Z$s,
and give a factor of $-1$ for each $Y$.  If we do this to the elements of
the stabilizer and get other elements of the stabilizer, this is a valid
fault-tolerant operation.  The seven-qubit code is an example of a code
for which this is true.

Another common bitwise operation is the $i$ phase
\begin{equation}
P = \pmatrix{ 1 & 0 \cr 0 & i}.
\end{equation}
On the basic operations $X$, $Y$, and $Z$ it acts as follows:
\begin{eqnarray}
P X P^\dagger = \pmatrix{ 1 & \ 0 \cr 0 & i}
\pmatrix{ 0 & -i \cr 1 & \ 0} = \pmatrix{ 0 & -i \cr i & \ 0} = & i Y \\
P Y P^\dagger = \pmatrix{ 1 & \ 0 \cr 0 & i}
\pmatrix{ 0 & i \cr 1 & 0} = \pmatrix{ \ 0 & i \cr i & \ 0} = & i X \\
P Z P^\dagger = \pmatrix{ 1 & \ 0 \cr 0 & i}
\pmatrix{ 1 & 0 \cr 0 & i} = \pmatrix{ 1 & \ 0 \cr 0 & -1} = & Z.
\end{eqnarray}
This switches $X$ and $Y$, but with extra factors of $i$, so there must
be a multiple of 4 $X$s and $Y$s for this to be a valid operation.  Again,
the seven-qubit code is an example of one where it is.  Note that a factor
of $i$ appears generically in any operation that switches $Y$ with $X$
or $Z$, because $Y^2 = -1$, while $X^2 = Z^2 = +1$.  The operations
in $N(\cal{G})$ actually permute $\sigma_X = X$, $\sigma_Z = Z$, and
$\sigma_Y = iY$, but for consistency with earlier publications I have
retained the notation of $X$, $Y$, and $Z$.  The most general single
qubit operation in $N(\cal{G})$ can be viewed as a rotation of the
Bloch sphere permuting the three coordinate axes.

We can also consider two-qubit operations, such as the controlled NOT.
Now we must consider transformations of the two involved blocks
combined.  The stabilizer group of the two blocks is $S \times S$, and
we must see how the basic operations $X \otimes I$, $Z \otimes I$,
$I \otimes X$, and $I \otimes Z$ transform under the proposed operation.
In fact, we will also need to know the transformation of $X \otimes Y$
and other such operators, but the transformation induced on $\cal{G}
\times \cal{G}$ is a group homomorphism, so we can determine the
images of everything from the images of the four elements listed above.

It is straightforward to show that the controlled NOT induces the following
transformation:
\begin{eqnarray}
X \otimes I & \rightarrow & X \otimes X \nonumber \\
Z \otimes I & \rightarrow & Z \otimes I \\
I \otimes X & \rightarrow & I \otimes X \nonumber \\
I \otimes Z & \rightarrow & Z \otimes Z. \nonumber
\end{eqnarray}
It is easy to see here how amplitudes are copied forwards and phases are
copied backwards.  The transformation laws for $R$, $P$, and CNOT are
also given in \cite{calderbank2}.

There are a number of basic gates in $N(\cal{G})$ beyond the ones given above.
As with the examples above, any gate can be characterized
by its transformation of the generators of $\cal{G}$ (or $\cal{G} \times
\cal{G}$ for two-qubit operations, and so on).  The primary constraint
that must be met is to preserve the algebraic properties of the operators.
In fact, there is a complete equivalence between the possible
gates and the automorphisms of $D_4$ (the group of products of $I$, $X$,
$Y$, and $Z$) or direct products of copies of $D_4$ (for multiple-qubit
gates) \cite{knillpc}.

Given any such automorphism, we first substitute $iY$ for $Y$ to get the
actual transformation.  Then we note that $|0\rangle$ is the ``encoded
zero'' for the ``code'' with stabilizer $\{I, Z\}$.  We know how $Z$
transforms under $U$, so $|0\rangle$ transforms to the state fixed
by $U Z U^\dagger$.  In addition, $|1\rangle = X |0\rangle$, so
$U |1\rangle = U X U^\dagger U |0\rangle$.  For instance, consider
the cyclic transformation 
\begin{equation}
T = X \rightarrow iY \rightarrow Z \rightarrow X.
\label{cyclic}
\end{equation}
Since $Z \rightarrow X$, 
\begin{equation}
|0\rangle \rightarrow 1/\sqrt{2}\ (|0\rangle + |1\rangle).
\end{equation}
Also, $X \rightarrow iY$, so 
\begin{equation}
|1\rangle \rightarrow i/\sqrt{2}\ Y(|0\rangle + |1\rangle)
= -i/\sqrt{2}\ (|0\rangle - |1\rangle).
\end{equation}
Thus, the matrix for $T$ is
\begin{equation}
T = \frac{1}{\sqrt{2}} \pmatrix{ 1 & -i \cr 1 & \ i}.
\end{equation}
We can perform a similar procedure to determine the matrix corresponding
to a multiple-qubit transformation.

The next question of interest is how much have we restricted our
computational power by restricting our attention to the normalizer
of $\cal{G}$?  Again, the normalizer of $\cal{G}$ is
exactly the group generated by the Hadamard transform $R$, the
phase $P$, and the controlled-NOT.  I will
prove this in section~\ref{measurements}.  Unfortunately, this
group alone is of only limited interest.  Knill~\cite{knillpc} has shown 
that a quantum computer using only operations from this group can
be simulated efficiently on a classical computer.\footnote{The
argument goes as follows: we start with an $n$-qubit state $|0\rangle$ 
which is the single state for the stabilizer code $\langle Z_1, \ldots,
Z_n \rangle$.  Each operation transforms the state and the stabilizer as 
above.  We can follow each transformation on a classical computer in
$O(n^2)$ steps.  A measurement picks at random one of the basis
kets in the codeword, which can also be chosen classically 
\cite{gottesman,cleve}.  This still leaves the question of partial
measurement of the full state, but the results
of section~\ref{measurements} show that this can also be classically
simulated.}  However, the addition of just the Toffoli gate 
to this group is sufficient to make the group universal \cite{shor2}.

\section{Measurements}
\label{measurements}

Now I will discuss what happens if we perform a measurement on a stabilizer
code.  Measuring individual qubits of an actual code is not of great
interest, but the results of this section will be quite helpful in 
determining what can be done by combining measurements and specific
fault-tolerant operations.

Now, using the method of DiVincenzo and Shor~\cite{divincenzo}, we
can measure any operator $A$ in ${\cal G}$.  There are three possible
relationships between $A$ and $S$.  First of all, $A$ could actually
be in $S$.  Then measuring $A$ tells us nothing about the state of
the system and does not change it at all.  The result of this measurement
will always be $+1$ for a valid codeword.  The second possibility is
for $A$ to commute with everything in $S$ but not to actually be in
$S$.  Then $A$ is equivalent to a nontrivial element of $N(S)/S$ and
measuring it will give us information about the state of the system.
This is usually inadvisable.

The third possibility, that $A$ anticommutes with something in $S$, is 
the most interesting.  In this case, we can choose the generators of
$S$ so that $A$ anticommutes with the first generator $M_1$ and commutes
with the remaining generators $M_2, \ldots, M_{n-k}$ (we can do this
since if generator $M_j$ anticommutes with $A$, we can replace it with
$M_1 M_j$, which commutes).  Then measuring $A$ does not disturb the
eigenvectors of $M_2$ through $M_{n-k}$, so those still fix the new state,
and are in the new stabilizer.  The eigenvectors of $M_1$ are disturbed,
however, and $M_1$ no longer fixes the states.  Measuring $A$ applies one
of the projection operators $P_+$ or $P_-$, where
\begin{equation}
P_{\pm} = \frac{1}{2} (I \pm A).
\end{equation}
Then $M_1^\dagger P_- M_1 = M_1^\dagger M_1 P_+ = P_+$, so if $|\psi\rangle$ is
some codeword,
\begin{equation}
M_1^\dagger P_- |\psi\rangle = M_1^\dagger P_- M_1 |\psi\rangle = 
P_+ |\psi\rangle.
\end{equation}
If the measurement result is $+1$, we do nothing else, and have 
thus applied $P_+$.  If the
measurement result is $-1$, apply $M_1^\dagger = M_1$, resulting in
the overall application of $P_+$.  Either way, $A$ fixes the new state.
This puts the system into the space
with stabilizer generated by $A, M_2, \ldots, M_{n-k}$.  From now on,
I will often say ``measure'' when I mean ``measure and correct for a
result of $-1$.''

Note that this construction works outside the framework of stabilizer
codes.  All we really need is a state $|\psi\rangle$, with $M|\psi\rangle
=|\psi\rangle$ for some unitary $M$.  Then, as above, we can perform
the projection $P_+$ for any operator $A$ satisfying $A^2=1$ and
$\{M, A\} = 0$.  Note that if $A$ is some local measurement,
either $|\psi\rangle$ is not entangled between the region affected by
$A$ and the region unaffected by it, or $M$ is a nonlocal operator.

We will want to know just where in the space a given state goes.  To
do this, look at the elements of $N(S)/S$.  If the
state starts in an eigenvector of $N$, it will also be an eigenvector
of $N' = MN$ for all $M \in S$.  After measuring $A$, the state will
no longer be an eigenvector of $N$ if $N$ anticommutes with $A$, but
it {\em will} still be an eigenvector of $M_1 N$, which commutes
with $A$.  Furthermore, the eigenvalue of $M_1 N$ stays the same.
Therefore, by measuring $A$ (and correcting the state if the
result is $-1$), we effectively transform the operator $N$ into $M_1 N$.
We could equally well say it is transformed to $M M_1 N$ instead,
where $M \in S$ commutes with $A$, but this will produce the same
transformation of the cosets of $N(S)/S$ to $N(S')/S'$ (where $S'$ is
the stabilizer after the measurement).  Of course, if $N$ commutes
with $A$, measuring $A$ leaves $N$ unchanged.

Let us see how all this works with a simple, but very useful, example.
Suppose we have two qubits, one in an arbitrary state $|\psi \rangle$,
the other initialized to $|0\rangle$.  The space of possible states
then has stabilizer $I \otimes Z$.  Suppose we perform a controlled-NOT
from the first qubit to the second.  This transforms the stabilizer to
$Z \otimes Z$.  Now let us measure the operator $I \otimes iY$ (we use
the factor of $i$ to ensure that the result is $\pm 1$).  This anticommutes
with $Z \otimes Z$, so if we get $+1$, we leave the result alone, and
if we get $-1$, we apply $Z \otimes Z$ to the state.  The new state is
in a $+1$-eigenstate of $I \otimes iY$, that is, $|\phi \rangle 
(|0\rangle + i |1\rangle)$.

How is $|\psi \rangle$ related to $|\phi \rangle$?  For the original
``code,'' $\overline{X} = X \otimes I$ and $\overline{Z} = Z \otimes I$.
After the CNOT, $\overline{X} = X \otimes X$ and $\overline{Z} = Z \otimes I$.
$X \otimes X$ does not commute with $I \otimes iY$, but the equivalent 
operator $Y \otimes Y = (X \otimes X) (Z \otimes Z)$ does.  $Z \otimes I$ 
does commute with $I \otimes iY$, so that stays the same.  Since the
second qubit is guaranteed to be in the $+1$ eigenstate of $iY$, we
might as well ignore it.  The effective $\overline{X}$ and $\overline{Z}$
operators for the first qubit are thus $-iY$ and $Z$ respectively.
This means we have transformed $\overline{X} \rightarrow -i \overline{X}
\overline{Z}$ and $\overline{Z} \rightarrow \overline{Z}$.  This is
the operation $P^\dagger$.

This example is simple enough that it is easy to check:
\begin{eqnarray}
|00\rangle & \rightarrow & |00\rangle = |0\rangle \frac{1}{2} \left[
(|0\rangle + i |1\rangle) + (|0\rangle - i |1\rangle)\right] \\
& \rightarrow & |0\rangle (|0\rangle \pm i |1\rangle) \\
& \rightarrow & |0\rangle (|0\rangle + i |1\rangle) \\
|10\rangle & \rightarrow & |11\rangle = |1\rangle \frac{i}{2} \left[
- (|0\rangle + i |1\rangle) + (|0\rangle - i |1\rangle) \right] \\
& \rightarrow & i |1\rangle (\mp |0\rangle - i|1\rangle) \\
& \rightarrow & \pm i |1\rangle (\mp |0\rangle \mp i|1\rangle) =
-i |1\rangle (|0\rangle + i |1\rangle).
\end{eqnarray}
Thus, ignoring the second qubit gives $|0\rangle \rightarrow |0\rangle$
and $|1\rangle \rightarrow -i|1\rangle$, which is $P^\dagger$.

This result is already quite interesting when coupled with the observation
that $P$ and CNOT suffice to produce $R$ as long as we can prepare and
measure states in the basis $|0\rangle \pm |1\rangle$ \cite{knill3}.
To do this we start out with the state $|\psi \rangle$ plus an ancilla
$|0\rangle + |1\rangle$.  Thus, the initial stabilizer is $I \otimes X$,
$\overline{X} = X \otimes I$, and $\overline{Z} = Z \otimes I$.  Apply
a CNOT from the second qubit to the first.  Now the stabilizer is
$X \otimes X$, $\overline{X} = X \otimes I$, and $\overline{Z} = Z \otimes Z$.
Apply $P$ to the second qubit, so the stabilizer is $X \otimes iY$,
$\overline{X} = X \otimes I$, and $\overline{Z} = Z \otimes Z$.  Measure
$I \otimes X$, performing $X \otimes iY$ if the result is $-1$.  This
produces $\overline{X} = X \otimes I$ and $\overline{Z} = iY \otimes X$,
so dropping the second qubit results in the transformation $Q$: 
$X \rightarrow X$, $Z \rightarrow iY$.  But $R = P Q^\dagger P$:
\begin{eqnarray}
X \rightarrow & iY \rightarrow\ Z\ \rightarrow & Z \\
Z \rightarrow &\ Z \rightarrow -iY \rightarrow & X
\end{eqnarray}
Coupled with the previous result, which derives $P$ from CNOT, this allows 
us to get any single qubit transformation in the normalizer of $\cal{G}$
provided we can perform a CNOT operation.

Another interesting application is to gain a new viewpoint on quantum
teleportation.  Suppose we have three qubits which start in the state
$|\psi\rangle (|00\rangle + |11\rangle)$.  The initial stabilizer is
$I \otimes X \otimes X$ and $I \otimes Z \otimes Z$, $\overline{X} = X 
\otimes I \otimes I$, and $\overline{Z} = Z \otimes I \otimes I$.
We assume the third qubit is far away, so we can do no operations
interacting it directly with the other two qubits.  We can, however,
perform operations on it conditioned on the result of measuring the
other qubits.  We begin by performing a CNOT from qubit one to two.
The stabilizer is now $I \otimes X \otimes X$ and $Z \otimes Z \otimes Z$,
$\overline{X} = X \otimes X \otimes I$, and $\overline{Z} = Z \otimes I
\otimes I$.  Measure $X$ for qubit one and discard qubit one.  If
the measurement result was $+1$, we leave the state alone; if it
was $-1$, we perform $Z$ on qubits two and three.  The stabilizer is
now $X \otimes X$, $\overline{X} = X \otimes I$ and $\overline{Z} =
Z \otimes Z$.  Now measure $Z$ for the new first qubit.  If the
result is $+1$, we leave the final qubit alone; if it is $-1$, we
apply $X$ to the last qubit.  This results in $\overline{X} = X$ and
$\overline{Z} = Z$, both acting on the last qubit.  We have succesfully
teleported the state $|\psi\rangle$.  The operations conditioned on
measurement results needed for teleportation arise here naturally as
the corrections to the stabilizer for alternate measurement results.
The formalism would have told us just as easily what operations were
necessary if we had begun with a different Bell state or a more complicated
entangled state (as long as it can still be described by a stabilizer).

I claimed before that products of $R$, $P$, and CNOT actually gave us all 
of the elements of $N(\cal{G})$, and I am now ready to prove that.  The
one-qubit operations correspond to the six automorphisms of $D_4$ given
by $R$, $P$, $Q$, $T$, $T^2$, and of course the identity.  We have
already seen that $Q = P^\dagger R P^\dagger$.  Also, $T = P Q^\dagger$, 
so all one-qubit operations are covered.

We can also perform all two-qubit operations.  Every automorphism of
$D_4 \times D_4$ can be produced by a composition of controlled NOT
and single-qubit operations.  For instance, take
\begin{eqnarray}
Z \otimes I & \rightarrow & X \otimes X \nonumber \\
I \otimes Z & \rightarrow & Z \otimes Z \\
X \otimes I & \rightarrow & iY \otimes X \nonumber\\
I \otimes X & \rightarrow & iZ \otimes Y. \nonumber
\end{eqnarray}
This permutation can be produced by performing the cyclic permutation
$X \rightarrow iY \rightarrow Z \rightarrow X$ on the first qubit and
a phase rotation $X \rightarrow iY$ on the second qubit, and
then performing a standard controlled NOT from the first qubit to the
second qubit.  It is straightforward to consider the other possibilities
and show that they too can be written using a CNOT and one-qubit gates.

I will show that the larger gates can be made this way by induction on
the number of qubits.  Suppose we know this to be true for all $n$-qubit
gates, and we have an $(n+1)$-qubit gate $U$.  On an arbitrary input state
$|0\rangle |\psi \rangle + |1\rangle |\phi \rangle$ (where $|\psi \rangle$ 
and $|\phi \rangle$ are $n$-qubit states), the output state will be
\begin{equation}
(|0\rangle |\psi_1 \rangle + |1\rangle |\psi_2 \rangle) +
(|0\rangle |\phi_1 \rangle + |1\rangle |\phi_2 \rangle).
\end{equation}
Suppose that under the applied transformation, $M = U (Z \otimes I \otimes
\cdots \otimes I) U^\dagger$ anticommutes with $Z \otimes I \otimes \cdots
\otimes I$.  If it does not, we can apply a one-qubit transformation and/or
rearrange qubits so that $M = X \otimes M'$, where $M'$ is an $n$-qubit
operation.  Suppose we apply $U$ to $|0\rangle |\psi\rangle$.  If we were 
then to measure $Z$ for the first qubit, we would get
either $0$, in which case the other qubits are in state $|\psi_1 \rangle$,
or $1$, in which case the remaining qubits are in state $|\psi_2 \rangle$.
The above analysis of measurements shows that $|\psi_1 \rangle$ and
$|\psi_2 \rangle$ are therefore related by the application of $M'$.
Define $U'$ by $U' |\psi \rangle = |\psi_1 \rangle$.  Then
\begin{equation}
U (|0\rangle |\psi \rangle) = (I + M) (|0\rangle \otimes U' |\psi \rangle).
\end{equation}

Let $N = U (X \otimes I \otimes \cdots \otimes I) U^\dagger$.  Again, we can
apply a one-qubit operation so that either $N = Z \otimes N'$ or 
$N = I \otimes N'$.  We can always put $M$ and $N$ in this form
simultaneously.  Then
\begin{eqnarray}
U (|1\rangle |\phi \rangle) & = & N U (|0\rangle |\phi \rangle) \\
& = & N (I + M) (|0\rangle \otimes U' |\phi \rangle) \\
& = & (I - M) N (|0\rangle \otimes U' |\phi \rangle) \\
& = & (I - M) (|0\rangle \otimes N' U' |\phi \rangle),
\end{eqnarray}
using the above form of $N$ and the fact that $\{M, N\} = 0$.

Now, $U'$ is an $n$-qubit operation, so we can build it out of $R$, $P$,
and CNOT.  To apply $U$, first apply $U'$ to the last $n$ qubits.
Now apply $N'$ to the last $n$ qubits conditioned on the first qubit being 
$1$.  We can do this with just a series of CNOTs and one-qubit operations.
Now apply a Hadamard transform to the first qubit.  This puts the system
in the state
\begin{equation}
(|0\rangle + |1\rangle) \otimes U' |\psi \rangle + 
(|0\rangle - |1\rangle) \otimes N' U' |\phi \rangle.
\end{equation}
Now, apply $M'$ to the last $n$ qubits conditioned on the first qubit.
Again, we can do this with just CNOTs and one-qubit operations.  This
leaves the system in the state
\begin{eqnarray}
& & |0\rangle \otimes U' |\psi \rangle + |1\rangle \otimes M' U' |\psi 
\rangle +
|0\rangle \otimes N' U' |\phi \rangle - |1\rangle \otimes M' N' U'
|\phi \rangle \\
& = & |0\rangle \otimes U' |\psi \rangle + M (|0\rangle \otimes U' |\psi 
\rangle) +
|0\rangle \otimes N' U' |\phi \rangle - M (|0\rangle \otimes N' U' |\phi
\rangle) \\
& = & (I + M) (|0\rangle \otimes U' |\psi \rangle) +
(I - M) (|0\rangle \otimes N' U' |\phi \rangle),
\end{eqnarray}
which we can recognize as the desired end state after applying $U$.

\section{Operations on CSS Codes}

In this section, I will finally begin to look at the problem of which
gates can be applied to specific codes.  One of the best classes of
codes for fault-tolerant computation are the Calderbank-Shor-Steane (CSS)
codes~\cite{calderbank1,steane1}, which are converted from certain 
classical codes.  These codes
have a stabilizer which can be written as the direct product of two
sectors, one of which is formed purely from $X$s and one formed just
from $Z$s.  These two sectors correspond to the two dual classical
codes that go into the construction of the code.

Shor \cite{shor2} showed that a punctured doubly-even self-dual CSS
code could be used for universal computation.  An example of such a
code is the seven-qubit code, whose stabilizer is given in 
Table~\ref{qubit7}.  
\begin{table}
\begin{tabular}{|l|ccccccc|}
$M_1$ & $X$ & $X$ & $X$ & $X$ & $I$ & $I$ & $I$ \\
$M_2$ & $X$ & $X$ & $I$ & $I$ & $X$ & $X$ & $I$ \\
$M_3$ & $X$ & $I$ & $X$ & $I$ & $X$ & $I$ & $X$ \\
$M_4$ & $Z$ & $Z$ & $Z$ & $Z$ & $I$ & $I$ & $I$ \\
$M_5$ & $Z$ & $Z$ & $I$ & $I$ & $Z$ & $Z$ & $I$ \\
$M_6$ & $Z$ & $I$ & $Z$ & $I$ & $Z$ & $I$ & $Z$ \\
\hline
$\overline{X}$ & $I$ & $I$ & $I$ & $I$ & $X$ & $X$ & $X$ \\
$\overline{Z}$ & $I$ & $I$ & $I$ & $I$ & $Z$ & $Z$ & $Z$ \\
\end{tabular}
\caption{The stabilizer and encoded $X$ and $Z$ for the seven-qubit code.}
\label{qubit7}
\end{table}
From the stabilizer, we can now understand why such codes allow the 
fault-tolerant implemention of the Hadamard rotation, the $\pi / 2$ 
rotation, and the controlled NOT.

The Hadamard rotation switches $X$ and $Z$.  For a CSS code, this is
a symmetry of the stabilizer if and only if the $X$ sector of the
stabilizer is the same as the $Z$ sector.  Therefore the two classical
codes had to be identical, and the quantum code must be derived from a
classical code that contains its own dual.  As we can see, this works for
the seven-qubit code.  In order to understand what the Hadamard rotation
does to the encoded states, we must look at what it does to the encoded 
$X$ and $Z$ operations.  For a punctured self-dual CSS code, the
$\overline{X}$ and $\overline{Z}$ operations can again be taken to be
the same, so the Hadamard rotation will just switch them.  It is therefore
an operation which switches encoded $X$ with encoded $Z$, and is thus
an encoded Hadamard rotation.

Similarly, for a self-dual code, the $\pi / 2$ rotation will convert
the $X$ generators into the product of all $Y$s.  This just converts
the $X$ generators into their product with the corresponding $Z$
generator, so this is a valid fault-tolerant operation, provided the
overall phase is correctly taken care of.  There is a factor of $i$
for each $X$, so there must be a multiple of 4 $X$s in each element of
the stabilizer for that to work out in general.  This will only be true
of a {\em doubly-even} CSS code, which gives us the other requirement
for Shor's methods.  Again, we can see that the seven-qubit code meets
this requirement.  Such a code will have $3\ {\rm mod}\ 4$ $X$s in the
$\overline{X}$ operation, so the bitwise $\pi /4$ converts $\overline{X}$
to $-i \overline{Y}$.  This is thus an encoded $- \pi / 2$ rotation.

Finally, we get to the controlled NOT.  This can be performed bitwise on
{\em any} CSS code.  We must look at its operation on $M \otimes I$ and
$I \otimes M$.  In the first case, if $M$ is an $X$ generator, it becomes
$M \otimes M$.  Since both the first and second blocks have the same
stabilizer, this is an element of $S \times S$.  If $M$ is a $Z$ generator,
$M \otimes I$ becomes $M \otimes I$ again.  Similarly, if $M$ is an $X$
generator, $I \otimes M$ becomes $I \otimes M$, and if $M$ is a $Z$
generator, $I \otimes M$ becomes $M \otimes M$, which is again in 
$S \times S$.  For an arbitrary CSS code, the $\overline{X_i}$ operators
are formed from the product of all $X$s and the $\overline{Z_i}$
operators are formed from the product of all $Z$s.  Therefore,
\begin{eqnarray}
\overline{X_i} \otimes I & \rightarrow & 
\overline{X_i} \otimes \overline{X_i} \nonumber \\
\overline{Z_i} \otimes I & \rightarrow & \overline{Z_i} \otimes I \\
I \otimes \overline{X_i} & \rightarrow & 
I \otimes \overline{X_i} \nonumber \\
I \otimes \overline{Z_i} & \rightarrow &
\overline{Z_i} \otimes \overline{Z_i}. \nonumber
\end{eqnarray}
Thus, the bitwise CNOT produces an encoded CNOT for every encoded qubit
in the block.

In fact, we can now easily prove that codes of the general CSS form are
the only codes for which bitwise CNOT is a valid fault-tolerant operation.
Let us take a generic element of the stabilizer and write it as $MN$, where
$M$ is the product of $X$s and $N$ is the product of $Z$s.  Then under
bitwise CNOT, $MN \otimes I \rightarrow MN \otimes M$, which implies
$M$ itself is an element of the stabilizer.  The stabilizer is a group,
so $N$ is also an element of the stabilizer.  Therefore, the
stabilizer breaks up into a sector made solely from $X$s and one made
solely from $Z$s, which means the code is of the CSS type.

\section{The Five-Qubit Code}
\label{five-qubit}

One code of particular interest is the five-qubit code
\cite{bennett,laflamme}, which is the
smallest possible code to correct a single error.  Until now, there were
no known fault-tolerant operations that could be performed on this
code except the simple encoded $X$ and encoded $Z$.  One presentation
\cite{calderbank2} of the five-qubit code is given in Table~\ref{qubit5}.
\begin{table}
\begin{tabular}{|l|ccccc|}
$M_1$ & $X$ & $Z$ & $Z$ & $X$ & $I$ \\
$M_2$ & $I$ & $X$ & $Z$ & $Z$ & $X$ \\
$M_3$ & $X$ & $I$ & $X$ & $Z$ & $Z$ \\
$M_4$ & $Z$ & $X$ & $I$ & $X$ & $Z$ \\
\hline
$\overline{X}$ & $X$ & $X$ & $X$ & $X$ & $X$ \\
$\overline{Z}$ & $Z$ & $Z$ & $Z$ & $Z$ & $Z$ \\
\end{tabular}
\caption{The stabilizer and encoded $X$ and $Z$ for the five-qubit code.}
\label{qubit5}
\end{table}
This presentation has the advantage of being cyclic, which simplifies
somewhat the analysis below.

This stabilizer is invariant under the transformation $T : X \rightarrow iY
\rightarrow Z \rightarrow X$ bitwise.  For instance,
\begin{equation}
M_1 = X \otimes Z \otimes Z \otimes X \otimes I \rightarrow
- Y \otimes X \otimes X \otimes Y \otimes I = M_3 M_4.
\end{equation}
By the cyclic property of the code, $M_2$ through $M_4$ also get transformed
into elements of the stabilizer, and this is a valid fault-tolerant operation.
It transforms
\begin{equation}
\overline{X} \rightarrow i \overline{Y} \rightarrow \overline{Z}.
\end{equation}
Therefore, this operation performed bitwise performs an encoded version of
itself.  Operations which have this property are particularly useful
because they are easy to apply to concatenated
codes~\cite{knill3,aharonov,knill2}.

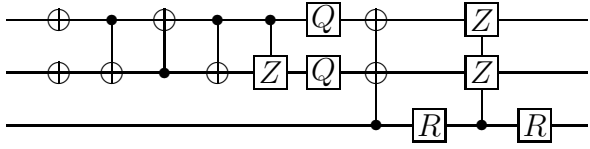
\begin{figure}
\begin{picture}(220,80)

\put(0,20){\line(1,0){154}}
\put(166,20){\line(1,0){28}}
\put(206,20){\line(1,0){14}}
\put(0,40){\line(1,0){94}}
\put(106,40){\line(1,0){8}}
\put(126,40){\line(1,0){48}}
\put(186,40){\line(1,0){34}}
\put(0,60){\line(1,0){114}}
\put(126,60){\line(1,0){48}}
\put(186,60){\line(1,0){34}}

\put(20,60){\circle{8}}
\put(20,56){\line(0,1){8}}
\put(20,40){\circle{8}}
\put(20,36){\line(0,1){8}}

\put(40,60){\line(0,-1){24}}
\put(40,60){\circle*{4}}
\put(40,40){\circle{8}}

\put(60,40){\line(0,1){24}}
\put(60,40){\circle*{4}}
\put(60,60){\circle{8}}

\put(80,60){\line(0,-1){24}}
\put(80,60){\circle*{4}}
\put(80,40){\circle{8}}

\put(100,60){\line(0,-1){14}}
\put(94,34){\framebox(12,12){$Z$}}
\put(100,60){\circle*{4}}

\put(114,54){\framebox(12,12){$Q$}}
\put(114,34){\framebox(12,12){$Q$}}

\put(140,20){\line(0,1){44}}
\put(140,20){\circle*{4}}
\put(140,40){\circle{8}}
\put(140,60){\circle{8}}

\put(154,14){\framebox(12,12){$R$}}

\put(180,20){\line(0,1){14}}
\put(180,46){\line(0,1){8}}
\put(180,20){\circle*{4}}
\put(174,34){\framebox(12,12){$Z$}}
\put(174,54){\framebox(12,12){$Z$}}

\put(194,14){\framebox(12,12){$R$}}

\end{picture}
\caption{Network to perform the $T_3$ gate.}
\label{buildT3}
\end{figure}

There is no nontrivial two-qubit operation in the normalizer of $\cal{G}$
that can be performed transversally on this code.  However, there is a
three-qubit transformation $T_3$ that leaves $S \times S \times S$ invariant:
\begin{eqnarray}
X \otimes I \otimes I & \rightarrow & iX \otimes Y \otimes Z \nonumber \\
Z \otimes I \otimes I & \rightarrow & iZ \otimes X \otimes Y \nonumber \\
I \otimes X \otimes I & \rightarrow & iY \otimes X \otimes Z \\
I \otimes Z \otimes I & \rightarrow & iX \otimes Z \otimes Y \nonumber \\
I \otimes I \otimes X & \rightarrow & X \otimes X \otimes X \nonumber \\
I \otimes I \otimes Z & \rightarrow & Z \otimes Z \otimes Z. \nonumber
\end{eqnarray}
On operators of the form $M \otimes I \otimes I$ or $I \otimes M \otimes I$,
this transformation applies cyclic transformations as above to the other
two slots.  Operators $I \otimes I \otimes M$ just become $M \otimes M \otimes
M$, which is clearly in $S \times S \times S$.  The matrix of $T_3$ is (up
to normalization)
\begin{equation}
T_3 = \pmatrix{
\ 1 & \ 0 & \ i & \ 0 & \ i & \ 0 & \ 1 & \ 0 \cr
\ 0 &  -1 & \ 0 & \ i & \ 0 & \ i & \ 0 &  -1 \cr
\ 0 & \ i & \ 0 & \ 1 & \ 0 &  -1 & \ 0 &  -i \cr
\ i & \ 0 &  -1 & \ 0 & \ 1 & \ 0 &  -i & \ 0 \cr
\ 0 & \ i & \ 0 &  -1 & \ 0 & \ 1 & \ 0 &  -i \cr
\ i & \ 0 & \ 1 & \ 0 &  -1 & \ 0 &  -i & \ 0 \cr
 -1 & \ 0 & \ i & \ 0 & \ i & \ 0 &  -1 & \ 0 \cr
\ 0 & \ 1 & \ 0 & \ i & \ 0 & \ i & \ 0 & \ 1 }.
\end{equation}
As with $T$, this operation performs itself on the encoded states.  A
possible network to produce this operation (based on the construction
in section~\ref{measurements}) is given in figure~\ref{buildT3}.

If we add in the possibility of measurements, this three-qubit operation
along with $T$ will allow us to perform any operation in the normalizer
of $\cal{G}$.  I will describe how to do this on unencoded qubits, and
since $T$ and $T_3$ bitwise just perform themselves, this will tell us
how to do the same operations on the encoded qubits.

To perform $P$, first prepare two ancilla qubits in the state $|00\rangle$
and use the data qubit as the third qubit.  The original
stabilizer is $Z \otimes I \otimes I$ and $I \otimes Z \otimes I$,
$\overline{X} = I \otimes I \otimes X$, and $\overline{Z} = I \otimes
I \otimes Z$.  Now apply $T_3$, so that the stabilizer is $iZ \otimes
X \otimes Y$ and $iX \otimes Z \otimes Y$, $\overline{X} = X \otimes
X \otimes X$, and $\overline{Z} = Z \otimes Z \otimes Z$.  Measure $Z$ for
the second and third qubits.  The resulting $\overline{X} = iY \otimes
I \otimes Z$ and $\overline{Z} = Z \otimes Z \otimes Z$.  Dropping
the last two qubits, we have $X \rightarrow iY$ and $Z \rightarrow Z$,
which is $P$.  Again, $Q = T^\dagger P$ and $R = P Q^\dagger P$, so we 
can perform any single qubit operation.

To get a two-qubit operation, prepare a third qubit in the state 
$|0\rangle$ and apply $T_3$.  This results in the stabilizer
$Z \otimes Z \otimes Z$, $\overline{X_1} = i X \otimes Y \otimes Z$,
$\overline{X_2} = iY \otimes X \otimes Z$, $\overline{Z_1} = i Z \otimes
X \otimes Y$, and $\overline{Z_2} = i X \otimes Z \otimes Y$.  Measure $X$
for the second qubit and throw it out.  This leaves the transformation
\begin{eqnarray}
X \otimes I & \rightarrow & i Y \otimes I \nonumber \\
I \otimes X & \rightarrow & i Y \otimes Z \\
Z \otimes I & \rightarrow & i Z \otimes Y \nonumber\\
I \otimes Z & \rightarrow & i Y \otimes X. \nonumber
\end{eqnarray}
This operation can be produced by applying $Q$ to the second qubit (switching
$Z$ and $iY$), then a CNOT from the second qubit to the first one, then
$P$ to the first qubit and $T^2$ to the second qubit.  Therefore, we can
also get a CNOT by performing this operation with the appropriate one-qubit
operations.  This allows us to perform any operation we desire in the
normalizer of $\cal{G}$.  Note that section~\ref{anycode} provides us
with another way to get these operations.  Having two methods available
broadens the choices for picking the most efficient implementations.

In order to perform universal computation on the five-qubit code, we must
know how to perform a Toffoli gate.  Shor \cite{shor2} gave a method for
producing a Toffoli gate that relied on the ability to perform the gate
\begin{equation}
|a\rangle |b\rangle |c\rangle \rightarrow (-1)^{a(b c)} |a\rangle
|b\rangle |c\rangle,
\label{toffoli}
\end{equation}
where $|a\rangle$ is either $|0 \ldots 0 \rangle$ or $|1 \ldots 1 \rangle$
and $|b\rangle$ and $|c\rangle$ are encoded $0$s or $1$s.  For the codes
Shor considered, this gate could be performed by applying it bitwise,
because the conditional sign could be applied bitwise.  All of the qubits
in the first block are either $0$ or $1$, so a controlled conditional
sign from the first block will produce a conditional sign on the second
two blocks whenever the first block is $1$.

For the five-qubit code, this gate is not quite as straightforward, but
is still not difficult.  To perform the two-qubit conditional sign gate 
on the five-qubit code, we need to perform a series of one- and three-qubit
gates and measurements.  However, if we perform each of these gates and
measurements conditional on the value of $a$, we have performed the
conditional sign gate on $|b\rangle |c\rangle$ if and only if the
first block is $1$.

From this, the rest of Shor's construction of the Toffoli gate carries over
straightforwardly.  It involves a number of measurements and operations from
the normalizer of $\cal{G}$.  We have already discussed how to do all of
those.  The one remaining operation that is necessary is
\begin{equation}
|a\rangle |d\rangle \rightarrow (-1)^{a d} |a\rangle |d\rangle,
\end{equation}
where $|d\rangle$ is an encoded state and $|a\rangle$ is again all $0$s
or all $1$s.  However, this is just $\overline{Z}$ applied to $|d\rangle$
conditioned on the value of $a$, which can be implemented with a single
two-qubit gate on each qubit in the block.  Therefore, we can perform
universal fault-tolerant computation on the five-qubit code.

Note that there was nothing particularly unique about the five-qubit code
that made the construction of the Toffoli gate possible.  The only property
we needed was the ability to perform a conditional sign gate.

\section{Gates for any Stabilizer Code}
\label{anycode}

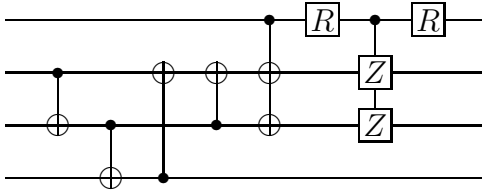
\begin{figure}
\begin{picture}(180,100)

\put(0,20){\line(1,0){180}}
\put(0,40){\line(1,0){134}}
\put(146,40){\line(1,0){34}}
\put(0,60){\line(1,0){134}}
\put(146,60){\line(1,0){34}}
\put(0,80){\line(1,0){114}}
\put(126,80){\line(1,0){28}}
\put(166,80){\line(1,0){14}}

\put(20,60){\line(0,-1){24}}
\put(20,60){\circle*{4}}
\put(20,40){\circle{8}}

\put(40,40){\line(0,-1){24}}
\put(40,40){\circle*{4}}
\put(40,20){\circle{8}}

\put(60,20){\line(0,1){44}}
\put(60,20){\circle*{4}}
\put(60,60){\circle{8}}

\put(80,40){\line(0,1){24}}
\put(80,40){\circle*{4}}
\put(80,60){\circle{8}}

\put(100,80){\line(0,-1){44}}
\put(100,80){\circle*{4}}
\put(100,60){\circle{8}}
\put(100,40){\circle{8}}

\put(114,74){\framebox(12,12){$R$}}

\put(140,80){\line(0,-1){14}}
\put(140,54){\line(0,-1){8}}
\put(140,80){\circle*{4}}
\put(134,54){\framebox(12,12){$Z$}}
\put(134,34){\framebox(12,12){$Z$}}

\put(154,74){\framebox(12,12){$R$}}

\end{picture}
\caption{Network to perform the four-qubit gate.}
\label{buildbig}
\end{figure}

Consider the following transformation:
\begin{eqnarray}
X\otimes I\otimes I\otimes I &\rightarrow & X\otimes X\otimes X\otimes I
\nonumber \\
I\otimes X\otimes I\otimes I &\rightarrow & I\otimes X\otimes X\otimes X 
\nonumber \\
I\otimes I\otimes X\otimes I &\rightarrow & X\otimes I\otimes X\otimes X 
\nonumber \\
I\otimes I\otimes I\otimes X &\rightarrow & X\otimes X\otimes I\otimes X \\
Z\otimes I\otimes I\otimes I &\rightarrow & Z\otimes Z\otimes Z\otimes I 
\nonumber \\
I\otimes Z\otimes I\otimes I &\rightarrow & I\otimes Z\otimes Z\otimes Z 
\nonumber \\
I\otimes I\otimes Z\otimes I &\rightarrow & Z\otimes I\otimes Z\otimes Z 
\nonumber \\
I\otimes I\otimes I\otimes Z &\rightarrow & Z\otimes Z\otimes I\otimes Z.
\nonumber
\end{eqnarray}
A possible gate array to perform this operation is given in 
figure~\ref{buildbig}.
This operation takes $M \otimes I \otimes I \otimes I$ to
$M \otimes M \otimes M \otimes I$, and cyclic permutations of this,
so if $M \in S$, the image of these operations is certainly in
$S \times S \times S \times S$.  This therefore is a valid transversal
operation on {\em any} stabilizer code.  The encoded operation it performs
is just itself.  There is a family of related operations for any even
number of qubits (the two-qubit case is trivial), but we only need to
concern ourselves with the four-qubit operation.

Suppose we have two data qubits.  Prepare the third and fourth qubits
in the state $|00\rangle$, apply the above transformation, and then 
measure $X$ for the third and fourth qubits.  The resulting transformation
on the first two qubits is then:
\begin{eqnarray}
X \otimes I & \rightarrow & X \otimes X \nonumber \\
I \otimes X & \rightarrow & I \otimes X \\
Z \otimes I & \rightarrow & Z \otimes I \nonumber \\
I \otimes Z & \rightarrow & Z \otimes Z. \nonumber
\end{eqnarray}
This is precisely the controlled NOT.  Since I showed in
section~\ref{measurements} that the CNOT was sufficient to get any
operation in $N(\cal{G})$, we can get any such operation for any
stabilizer code!  In fact, using the Toffoli gate construction from 
section~\ref{five-qubit}, we can perform universal computation.

Actually, this only gives universal computation for codes encoding a single
qubit in a block, since if a block encodes multiple qubits, this operation
performs the CNOT between corresponding encoded qubits in different
blocks.  To actually get universal computation, we
will want to perform operations between qubits encoded in the same
block.  To do this, we need a few more tools, which will be presented
in the next section.  I will also consider a few more examples where
we have tools beyond the ones available for any code.

\section{Distance 2 Codes}

There is a large class of distance 2 codes with a very simple form.  The
stabilizer for these codes has just two generators, one a product of all
$X$s and one a product of all $Z$s.  The total number of qubits $n$ must
be even.  These codes encode $n-2$ qubits, and therefore serve as a
good model for block codes encoding multiple qubits.  While these 
distance 2 codes cannot actually correct a general error, they may be 
useful in their own right nonetheless.  A
distance 2 code can be used for error detection \cite{vaidman}.  If we 
encode our computer using distance 2 codes, we will not be able to fix any
errors that occur, but we will know if an error has invalidated our 
calculation.  A better potential use of distance 2 codes is to fix located
errors \cite{grassl}.  Suppose the dominant error source in our hardware
comes from qubits leaving the normal computational space.  In principle,
without any coding, we can detect not only that this has happened, but
in which qubit it has occurred.  We can then use this information in
conjunction with a distance 2 code to correct the state, as with a usual
quantum error-correcting code.  A final possible use of distance 2 codes
is to concatenate them to produce codes that can correct multiple errors.
Since the limiting factor in the computational threshold for concatenated
codes is the time to do error correction, this offers potentially a great
advantage.  However, there is a significant complication in this program,
since the codes given here encode more than one qubit, which complicates
the concatenation procedure.

Because of the simple structure of these distance 2 codes, we can immediately
see a number of possible fault-tolerant operations.  The bitwise Hadamard
rotation and the bitwise CNOT are both permissible.  If the total number
of qubits is a multiple of 4, the $P$ gate and the other single
qubit operations are allowed, as well.  What is less clear is how these
various operations affect the encoded data.

The $\overline{X_i}$ operators for these codes are $X_1 X_{i+1}$, where
$i$ runs from 1 to $n-2$.  The $\overline{Z_i}$ operators are
$Z_{i+1} Z_n$.  Therefore, swapping the $(i+1)$th qubit with the $(j+1)$th
qubit will swap the $i$th encoded qubit with the $j$th encoded qubit.
Swapping two qubits in a block is not a transversal operation, but if
performed carefully, it can still be done fault-tolerantly.  One
advantage of the swap operation is that any errors in one qubit will
not propagate to the other, since they are swapped as well.  However,
applying the swap directly to the two qubits allows the possibility of
an error in the swap gate itself producing errors in both qubits.  We
can circumvent this by introducing a third ancilla qubit.  Suppose we
wish to swap A and B, which are in spots 1 and 2, using ancilla C, in
spot 3.  First swap the qubits in spots 1 and 3, then 1 and 2, and
finally 2 and 3.  Then A ends up in spot 2, B ends up in spot 1, and
C ends up in spot 3, but A and B have never interacted directly.  We
would need two swap gates to go wrong in order to introduce errors
to both A and B.  Note that while the state C does not matter, it should
not be something important, since it is exposed to error from all three
swap gates.  Also note that we should perform error correction before
interacting this block with another block, since errors could then spread
between corresponding qubits, which have changed.

The action of the CNOT is simple.  As for other CSS codes, it just produces
a CNOT between all of the encoded qubits in the first block with the
corresponding encoded qubits in the second block.  The Hadamard rotation
converts $\overline{X_i}$ to $Z_1 Z_{i+1}$, which is equivalent (via
multiplication by $M_2$) to $Z_2 \ldots Z_i Z_{i+2} \ldots Z_n$.  This
is equal to $\overline{Z_1} \ldots \overline{Z_{i-1}} \overline{Z_{i+1}}
\ldots \overline{Z_{n-2}}$.  Similarly, $\overline{Z_i}$ becomes
$\overline{X_1} \ldots \overline{X_{i-1}} \overline{X_{i+1}} \ldots
\overline{X_{n-2}}$.  For instance, for the smallest case, $n=4$,
\begin{eqnarray}
\overline{X_1} & \rightarrow & \overline{Z_2} \nonumber \\
\overline{Z_1} & \rightarrow & \overline{X_2} \\
\overline{X_2} & \rightarrow & \overline{Z_1} \nonumber \\
\overline{Z_2} & \rightarrow & \overline{X_1}. \nonumber
\end{eqnarray}
The Hadamard rotation for $n=4$ performs a Hadamard rotation on each 
encoded qubit and simultaneously switches them.  For larger $n$, it
performs the Hadamard rotation on each qubit, and performs a variation
of the class of codes discussed in section~\ref{anycode}.

For $n=4$, the $P$ gate acts as follows:
\begin{eqnarray}
\overline{X_1} & \rightarrow & -Y_1 Y_2 = -\overline{X_1} \overline{Z_2} 
\nonumber \\
\overline{X_2} & \rightarrow & -Y_1 Y_3 = -\overline{X_2} \overline{Z_1} \\
\overline{Z_1} & \rightarrow & \overline{Z_1} \nonumber \\
\overline{Z_2} & \rightarrow & \overline{Z_2}. \nonumber
\end{eqnarray}
A consideration of two-qubit gates allows us to identify this as a variant
of the conditional sign gate.  Specifically, this gate gives a sign of
$-1$ unless both qubits are $|0\rangle$.

When we allow measurement, a trick becomes available that is useful for
any multiple-qubit block code.  Given one data qubit, prepare a second
ancilla qubit in the state $|0\rangle + |1\rangle$, then apply a CNOT
from the second qubit to the first qubit and measure $Z$ for the first qubit.
The initial stabilizer is $I \otimes X$; after the CNOT it is $X \otimes X$.
Therefore the full operation takes $X \otimes I$ to $I \otimes X$ and
$Z \otimes I$ to $Z \otimes Z$.  We can discard the first qubit and the
second qubit is in the initial data state.  However, if we prepare the
ancilla in the state $|0\rangle$, then apply a CNOT, the original state
is unaffected.  Therefore, by preparing a block with all but the $j$th
encoded qubit in the state $|0\rangle$, and with the $j$th encoded qubit
in the state $|0\rangle + |1\rangle$, then applying a CNOT from the new
block to a data block and measuring the $j$th encoded qubit in the data
block, we can switch the $j$th encoded qubit out of the data block and
into the new, otherwise empty block.

This trick enables us to perform arbitrary operations on qubits from 
the same block for the distance 2 codes.
We switch the qubits of interest into blocks of their own, use swap
operations to move them into corresponding spots, then perform whole
block operations to interact them.  Then we can swap them back and
switch them back into place in their original blocks.

The step that is missing for arbitrary stabilizer codes is the ability
to move individual encoded qubits to different places within a block.
Since the gate in section~\ref{anycode} gives us a block CNOT, we can
perform the switching operation into an empty block.  By using switching
and whole block operations, we can perform an arbitrary one-qubit 
operation on any single encoded qubit within a block.  The only
remaining operation necessary is the ability to swap an encoded qubit
from the $i$th place to the $j$th place.  We can do this using
quantum teleportation.  All that is required is an otherwise empty
block with the $i$th and $j$th encoded qubits in the entangled state
$|00\rangle + |11\rangle$.  Then we need only perform single-qubit
operations and a CNOT between the qubits in the $i$th places, both
of which we can do.  To prepare the entangled state, we simply start
with the $+1$-eigenstate of $\overline{Z_i}$ and $\overline{Z_j}$,
then measure the eigenvalue of $\overline{X_i} \overline{X_j}$ (and
correct if the result is $-1$).  This is just an operator in ${\cal G}$,
so we know how to do this.  The state stays in an eigenvector of
$\overline{Z_i} \overline{Z_j}$, which commutes with $\overline{X_i}
\overline{X_j}$, so the result will be the desired encoded Bell state.
We can then teleport the $i$th qubit in one otherwise empty block to
the $j$th qubit in the block originally containing the entangled state.
This was all we needed to allow universal computation on any stabilizer
code.

\section{The 8 Qubit Code}

There is a code correcting one error encoding 3 qubits in 8 qubits
\cite{gottesman,calderbank2,steane2}.  The stabilizer is given in
table~\ref{qubit8}.
\begin{table}
\begin{tabular}{|l|cccccccc|}
$M_1$ & $X$ & $X$ & $X$ & $X$ & $X$ & $X$ & $X$ & $X$ \\
$M_2$ & $Z$ & $Z$ & $Z$ & $Z$ & $Z$ & $Z$ & $Z$ & $Z$ \\
$M_3$ & $X$ & $I$ & $X$ & $I$ & $Z$ & $Y$ & $Z$ & $Y$ \\
$M_4$ & $X$ & $I$ & $Y$ & $Z$ & $X$ & $I$ & $Y$ & $Z$ \\
$M_5$ & $X$ & $Z$ & $I$ & $Y$ & $I$ & $Y$ & $X$ & $Z$ \\
\hline
$\overline{X_1}$ & $X$ & $X$ & $I$ & $I$ & $I$ & $Z$ & $I$ & $Z$ \\
$\overline{X_2}$ & $X$ & $I$ & $X$ & $Z$ & $I$ & $I$ & $Z$ & $I$ \\
$\overline{X_3}$ & $X$ & $I$ & $I$ & $Z$ & $X$ & $Z$ & $I$ & $I$ \\
$\overline{Z_1}$ & $I$ & $Z$ & $I$ & $Z$ & $I$ & $Z$ & $I$ & $Z$ \\
$\overline{Z_2}$ & $I$ & $I$ & $Z$ & $Z$ & $I$ & $I$ & $Z$ & $Z$ \\
$\overline{Z_3}$ & $I$ & $I$ & $I$ & $I$ & $Z$ & $Z$ & $Z$ & $Z$ \\
\end{tabular}
\caption{The stabilizer and encoded $X$s and $Z$s for the eight-qubit code.}
\label{qubit8}
\end{table}
There are no transversal operations that leave this stabilizer fixed.
However, when we allow swaps between the constituent qubits, a number
of possibilities become available.

One possible operation is to swap the first four qubits with the second
four qubits.  This leaves $M_1$, $M_2$, and $M_4$ unchanged.  $M_3$ becomes
instead $M_1 M_2 M_3$, and $M_5$ becomes $M_1 M_5$.  On the encoded qubits,
this induces the transformation
\begin{eqnarray}
X \otimes I \otimes I & \rightarrow & X \otimes I \otimes Z \nonumber \\
I \otimes X \otimes I & \rightarrow & I \otimes X \otimes I \nonumber \\
I \otimes I \otimes X & \rightarrow & Z \otimes I \otimes X \\
Z \otimes I \otimes I & \rightarrow & Z \otimes I \otimes I \nonumber \\
I \otimes Z \otimes I & \rightarrow & I \otimes Z \otimes I \nonumber \\
I \otimes I \otimes Z & \rightarrow & I \otimes I \otimes Z. \nonumber
\end{eqnarray}
This is just a conditional sign on the first and third qubits, with
the second encoded qubit unaffected.  Through single-qubit transformations,
we can convert this to a controlled NOT, and using this perform a swap 
between the first and third encoded positions.

Another operation is to swap qubits one and two with three and four
and qubits five and six with seven and eight.  This leaves $M_1$, $M_2$,
and $M_3$ unchanged, and converts $M_4$ to $M_2 M_4$ and $M_5$ to $M_1 M_5$.
On the encoded qubits, it induces the transformation
\begin{eqnarray}
X \otimes I \otimes I & \rightarrow & X \otimes Z \otimes Z \nonumber \\
I \otimes X \otimes I & \rightarrow & Z \otimes X \otimes Z \nonumber \\
I \otimes I \otimes X & \rightarrow & Z \otimes Z \otimes X \\
Z \otimes I \otimes I & \rightarrow & Z \otimes I \otimes I \nonumber \\
I \otimes Z \otimes I & \rightarrow & I \otimes Z \otimes I \nonumber \\
I \otimes I \otimes Z & \rightarrow & I \otimes I \otimes Z. \nonumber
\end{eqnarray}
We could also switch the odd numbered qubits with the even numbered
qubits.  That leaves $M_1$ and $M_2$ unchanged, while turning $M_3$
into $M_1 M_3$, $M_4$ into $M_1 M_4$, and $M_5$ into $M_1 M_2 M_5$.
On the encoded qubits it induces
\begin{eqnarray}
X \otimes I \otimes I & \rightarrow & X \otimes I \otimes Z \nonumber \\
I \otimes X \otimes I & \rightarrow & I \otimes X \otimes Z \nonumber \\
I \otimes I \otimes X & \rightarrow & Z \otimes Z \otimes X \\
Z \otimes I \otimes I & \rightarrow & Z \otimes I \otimes I \nonumber \\
I \otimes Z \otimes I & \rightarrow & I \otimes Z \otimes I \nonumber \\
I \otimes I \otimes Z & \rightarrow & I \otimes I \otimes Z. \nonumber
\end{eqnarray}
This is just a conditional sign between the first and third places
followed by a conditional sign between the second and third places.
Combined with the first operation, it gives us a conditional sign
between the second and third places, which we can again convert to
a swap between the second and third encoded positions.  This allows
us to swap any two encoded qubits in the block, which is sufficient
to give us universal computation.

In this case, the symmetries of the code naturally became allowed
transformations of the stabilizer.  This is likely to hold true in
many other cases as well.  As with the five-qubit code, we also
had available a universal protocol for swapping encoded qubits,
but multiple methods again allow us more freedom in choosing
efficient methods.

\section{Summary and Discussion}

I have presented a general theory for understanding when it is possible to
apply a given operation transversally to a given quantum error-correcting
code, and for understanding the results of making a measurement
on a stabilizer code.  These results clarify the advantages of the 
doubly-even self-dual CSS codes used by Shor~\cite{shor2}.  They
also provide protocols for performing universal computation on
any stabilizer code.  In many cases, the protocols described
here call for a number of steps to perform most simple operations,
so more efficient protocols for specific codes are desirable,
and I expect the methods described in this paper will be quite
helpful when searching for these protocols.

Efficient use of space is also important.  Existing methods of
fault-tolerant computation use space very inefficiently, and
being able to use more efficient codes (such as those encoding
multiple qubits in a block) could be very helpful in reducing the
space requirements.

This work was supported in part by the U.S. Department of Energy under 
Grant No. DE-FG03-92-ER40701 and by DARPA under Grant No. DAAH04-96-1-0386
administered by the Army Research Office.  I would
like to thank John Preskill, Manny Knill, Richard Cleve, and David
DiVincenzo for helpful discussions.


\begin{thebibliography}{99}

\bibitem{shor1} P.~W.~Shor, Phys. Rev.~A {\bf 52}, 2493 (1995).

\bibitem{calderbank1} A.~R.~Calderbank and P.~W.~Shor, 
Phys. Rev.~A {\bf 54}, 1098 (1996).

\bibitem{steane1} A.~Steane, ``Multiple Particle Interference and
Quantum Error Correction,'' quant-ph/9601029, 1996.

\bibitem{gottesman} D.~Gottesman, Phys. Rev.~A {\bf 54}, 1862 (1996).

\bibitem{calderbank2} A.~R.~Calderbank, E.~M.~Rains, P.~W.~Shor, and
N.~J.~A.~Sloane, ``Quantum Error Correction and Orthogonal Geometry,''
quant-ph/9605005, 1996.

\bibitem{calderbank3} A.~R.~Calderbank, E.~M.~Rains, P.~W.~Shor, and
N.~J.~A.~Sloane, ``Quantum Error Correction Via Codes Over
GF(4),``quant-ph/9608006, 1996.

\bibitem{shor2} P.~W.~Shor, ``Fault-Tolerant Quantum Computation,''
quant-ph/9608012, 1996.

\bibitem{knill3} E.~Knill, R.~Laflamme, and W.~Zurek, ``Accuracy Threshold
for Quantum Computation,'' quant-ph/9610011, 1996.

\bibitem{aharonov} D.~Aharonov and M.~Ben-Or, ``Fault-Tolerant Quantum
Computation With Constant Error,'' quant-ph/9611025, 1996.

\bibitem{divincenzo} D.~DiVincenzo and P.~W.~Shor, ``Fault-Tolerant
Error Correction with Efficient Quantum Codes,'' quant-ph/9605031, 1996.

\bibitem{bennett} C.~H.~Bennett, D.~P.~DiVincenzo, J.~A.~Smolin, and
W.~K.~Wootters, Phys. Rev.~A {\bf 54}, 3824 (1996).

\bibitem{knill4} E.~Knill, ``Group Representations, Error Bases and
Quantum Codes,'' quant-ph/9608049, 1996.

\bibitem{knillpc} E.~Knill, personal communication.

\bibitem{cleve} R.~Cleve and D.~Gottesman, ``Efficient Computations of
Encodings for Quantum Error Correction,'' quant-ph/9607030, 1996.

\bibitem{laflamme} R.~Laflamme, C.~Miquel, J.~P.~Paz, and W.~H.~Zurek, 
``Perfect Quantum Error Correction Code,'' quant-ph/9602019, 1996.

\bibitem{knill2} E.~Knill and R.~Laflamme, ``Concatenated Quantum Codes,''
quant-ph/9608012, 1996.

\bibitem{vaidman} L.~Vaidman, L.~Goldenberg, and S.~Wiesner, Phys. Rev.~A
{\bf 54}, 1745R (1996).

\bibitem{grassl} M.~Grassl, T.~Beth, and T.~Pellizzari, ``Codes for the
Quantum Erasure Channel,'' quant-ph/9610042, 1996.

\bibitem{steane2} A.~Steane, Phys. Rev.~A {\bf 54}, 4741 (1996).

\end{thebibliography}
\end{document}